\documentclass[10pt, a4paper]{article}
\usepackage{xurl}
\usepackage[a4paper, total={6in, 8in}]{geometry}
\usepackage{abstract}

\usepackage{lipsum}
\usepackage{amsmath}
\usepackage{algorithm}
\usepackage[noend]{algpseudocode}
\usepackage{tikz}
\usepackage{array,graphicx}
\usepackage [latin1]{inputenc}

\usepackage{verbatim}
\usepackage{amssymb}
\usepackage{algorithm}
\usepackage[noend]{algpseudocode}

\usepackage{tikz,pgfplots}
\usepackage{array}
\usetikzlibrary{spy}
\usepackage{amsfonts}
\usepackage{latexsym}
\usepackage{amsmath}
\usepackage{amssymb}
\usepackage{amscd}
\usepackage{amsthm}
\usepackage{hyperref}
\usepackage{multirow}
\usepackage{array,graphicx}
 \usepackage{float}
\usepackage{setspace}
\usepackage{caption}
\usepackage{amsthm}
\usepackage{authblk}
\usepackage[shortlabels]{enumitem}

\usetikzlibrary{decorations.pathreplacing}
\tikzstyle{altfill}=[fill=white]
\pgfmathsetmacro{\sixvspacing}{0.36}
\pgfmathsetmacro{\topgspacing}{0.15}
\pgfmathsetmacro{\btmgspacing}{0.15}

\tikzstyle{smallgraph}=[baseline=-1.5ex,
  every node/.style={circle,draw,fill=black!30,inner sep=0pt,minimum width=4pt},
  every path/.append style={semithick}
]
\tikzstyle{largegraph}=[thick,scale=0.42,
every node/.style={circle, draw, fill=black!100,inner sep=0pt, minimum width=4pt}
]
\usepackage[T1]{fontenc} % Font encoding: T1r

\usetikzlibrary{decorations.pathreplacing}
\usepackage{tikz,pgfplots}

\pgfplotsset{compat=1.3}
\usepackage{graphicx} % Allows including images

\usepackage{hyperref}
\hypersetup{colorlinks=true, urlcolor=blue, linkcolor=blue, citecolor=blue}

\title{Simplifications to Guide Monte Carlo Tree Search in Combinatorial Games}

\author[1]{Michael Haythorpe}
\author[1]{Alex Newcombe}
\author[2]{Damian O'Dea}
\affil[1]{College of Science and Engineering, Flinders University, South Australia}
\affil[2]{Defence Science and Technology Group, Edinburgh, South Australia}
\begin{document}
%%%%%%%%%%%%
% Abstract %
%%%%%%%%%%%%

    \maketitle
    \begin{abstract}
     We examine a type of modified Monte Carlo Tree Search (MCTS) for strategising in combinatorial games. The modifications are derived by analysing simplified strategies and simplified versions of the underlying game and then using the results to construct an ensemble-type strategy. We present some instances where relative algorithm performance can be predicted from the results in the simplifications, making the approach useful as a heuristic for developing strategies in highly complex games, especially when simulation-type strategies and comparative analyses are largely intractable. 

    \end{abstract}

%%%%%%%%%%%
% Article %
%%%%%%%%%%%

\section{Introduction} \label{sec-intro}
Extremely complex games have many actions available at each turn and many turns before the game ends, and this often makes a meaningful simulation and comparison of strategies extremely difficult. One potential approach comes from the field of general game playing AI, which attempts to transfer strategic knowledge between games with similar underlying style of play \cite{piette}. Such methods have been successful for the game Go \cite{assayang} in which AI models were trained on a fixed-size board and then successfully applied to varying-sized boards. However, this type of knowledge transfer is not always successful; for example, \cite{raghu} observed instances of combinatorial games where deep reinforcement learning models trained and tested in slightly different versions resulted in a significant reduction in performance. Due to the disparity in previous results, it is interesting to examine further the types of scenarios where successful knowledge transfer can occur. In this paper, we approach this by examining sets of games that are closely related to each other and also gradually changing in complexity. By examining strategies in a range of games in the set, we can identify whether strategy performance is retained for the whole set. Of course, such relationships are dependent on the games and the types of strategies being used, and in this work we focus on Monte Carlo Tree Search (MCTS) along with a type of simplified strategy which we denote a micro-strategy.

When a knowledge transfer is identified as being possible, we utilise it to construct a metaheuristic. To do this, we describe a general method for augmenting a strategy in a complex combinatorial game by using strategies from simplified/related versions of the game. In our case, the result is a modified MCTS which is an improvement over a standard MCTS whenever the corresponding knowledge transfer was successful. The idea of utilising simplified strategies to augment a more complex MCTS comes from the field of combinatorial optimisation where, in some cases, simple properties that are not specifically related to the objective provide benefits to the optimisation process, for example, in the Hamiltonian Cycle Problem \cite{ejov}.

In Section \ref{sec-exp} we identify several types of combinatorial Maker-Breaker games for which the above approach is successful and the resulting algorithm provides an improvement over standard MCTS. The Maker-Breaker type games are a useful test bed for our investigations as there are many possibilities for altering the game in a minor way beyond, for example, changing the board size. In general, the types of games that are interesting for this approach are games whose parameters can be incremented such that each increment produces a new game which is closely related to the last. Usually it is not possible to prove ``closeness'' however intuitive reasoning is enough to warrant this type of experimental investigation.

In the remainder of this Section we introduce Maker Breaker games, Monte Carlo Tree Search and a type of strategy which is utilised in this approach, denoted Micro-strategies. Then, in Section \ref{sec-exp} we apply the approach to three types of game simplification, and some of the tested instances produce positive results, which we discuss further. 

\subsection{Maker Breaker Games} \label{sec:mbgames}
The combinatorial games utilised in the present work are versions of Maker-Breaker (MB) games. The original definition of an MB game \cite{erdos} is: Given a graph $G$ (the board) and a collection $W$ of subsets of the vertices (the winning sets), two players named {\em Maker} and {\em Breaker} take turns marking a vertex of $G$. If Maker manages to mark all vertices from a set in $W$ then Maker wins, otherwise, Breaker wins. The winning sets of an MB game are commonly given implicitly as any set of vertices from the game board whose induced subgraph is isomorphic to a given fixed graph, for example, a cycle of a given length. We shall also use this convention and the winning sets will be produced from all paths of a given length, or else, all vertices which form a dominating set \footnote[1]{A dominating set of a graph is a subset $S$ of vertices such that every vertex is either in $S$ or adjacent to a member of $S$.}.

MB games are versatile in that many alternative versions can be created by adding or modifying the rules. In particular, we are interested in games that can be altered in {\em minor} ways, in the hope that doing so retains many characteristics of the original game. In these cases, the performance of strategies may be somewhat retained across the minor changes. Some examples of modifications which may retain characteristics include altering the board, the winning sets and the turn progression of the game. In order to represent game changes and simplifications, we will use sets of MB games. Ideally, a set contains small/simple games which then vary in a minor way, creating new members of the set, eventually scaling up to very large/complex games. Some sets with these properties are described in detail later in Section \ref{sec-typesofsimplifications}, but there are many possible ways to produce such sets. Note that in an application, there is usually a single complex game of interest, and there may be several different ways of simplifying this game, each producing a different set of games for investigation. 

\subsection{Monte Carlo Tree Search in Games}
Monte Carlo Tree Search (MCTS) is a popular algorithm to search the state and action space of a decision process. In games, it can be effective as a standalone strategy, but also many of the prominent game-playing algorithms use an auxiliary MCTS to benefit some larger AI process. 

The game-tree of a game is a rooted, directed tree, whose root vertex corresponds to the initial game state. A directed edge is created for each possible move and a new vertex is created representing the corresponding updated game state when that move is taken. If different move sequences lead to the same game-state, then there are multiple vertices in the game-tree which correspond to the same game-state. MCTS is a heuristic search algorithm which can be applied to a game-tree and provides a simulation based approach to strategising in the game. The algorithm combines the exploitation of previously found good moves and the exploration of new, not yet considered moves. MCTS proceeds by iteratively constructing a subtree of the game tree and records information about the vertices/game states in the current subtree. Initially, a root vertex is added which corresponds to the current game state. The search then progresses in four phases:
\begin{itemize}\label{alg1}
\item Selection: Traverse from the root to a leaf in the current subtree. The path taken is determined by an evaluation function whose purpose is to weight the (so far) best sequence of moves.

\item Expansion: Add nodes below the leaf that correspond to the next possible set of moves. Choose the next move, represented by the vertex $v$.

\item Rollout: Play the game out from this state in a simple fashion (usually randomly) and record information about the each game result.

\item Back propagate: Update the information attached to the nodes on the path from the root to vertex $v$.
\end{itemize}

An evaluation function is needed to value the moves and usually, the true value of a move is unknown and an approximation is used instead. A common approximation value of a move during the selection phase is the upper confidence bound applied to trees (UCT). Each iteration of MCTS completes a selection path from the root to a leaf corresponding to a possible game end-state and it is a winning path if the player of interest has won. The UCT of a vertex $v$ in the current subtree is then 
\begin{equation}
\text{UCT}(v) \equiv \frac{w}{n} + c\sqrt{\frac{\log N}{n}}
\end{equation}
Where $w$ is the number of winning paths found containing $v$, $n$ is the number of selection paths containing $v$ and $N$ is the total number of selection paths. The left fraction is large for vertices which have commonly led to wins, while the right fraction is large for vertices that have not been explored often. Thus, the $c$ parameter provides control over the exploitation versus exploration aspect of MCTS. During the selection and expansion phases, there can be many moves that are assigned an equivalent best value, and traditional MCTS picks one of these at random. A modified MCTS uses some auxiliary algorithm to weight the choice, which may come from alternative evaluation functions or some other heuristic process. In particular, our implementation uses various other strategies, which are discussed in the next section, to weight both the selection phase and the gameplay during the rollout phase.

\subsection{Micro-Strategies for Weighted MCTS}\label{sec-ms}
Modified MCTS uses an auxiliary algorithm during the rollout and/or selection phases, however, to still benefit from the Monte Carlo aspect, auxiliary algorithms must be relatively computationally efficient. In an extreme case, it would be difficult to evaluate an NP-hard problem at each visited node during the selection/rollout phases of every iteration. The proposed auxiliary algorithms provide a weight for each possible move in a given game state, that is, they are strategy functions in their own right which we call {\em micro-strategies}. Micro-strategies provide a fast, albeit usually naive way to play, for example, counting the number of game pieces remaining after a move, or calculating the immediate increase in game score. In this context, a uniform random strategy is also a micro-strategy which weights each available move equally. Whenever MCTS makes a move randomly, it calls a uniform random micro-strategy, which may be replaced with any different micro-strategy, some of which may provide improvements to the overall performance of the MCTS. 

In an MB game a partially played game is represented as the underlying graph (game-board) with its vertices coloured depending on which player has marked them, including a colour for unmarked. Many micro-strategies can be created by analysing properties of the graph or coloured graph. In order to utilise micro-strategies for our approach, they must be applicable to any of the games from the sets of MB games defined in Section \ref{sec:mbgames}. To ensure this, we focus on micro-strategies which are well defined for any MB game, as opposed to specific strategies only usable for one particular game. Some examples of micro-strategies used in the experiments of Section \ref{sec-exp} include:

\begin{itemize}
    \item Choosing vertices with the large/small degree (either coloured or uncoloured degree)
    \item Choosing vertices with graph distance which is near/far-away from the previous moves
    \item Choosing vertices belonging to highly/lowly connected components in the graph 
    \item Choosing vertices belonging to many/few winning sets.
\end{itemize} 

As mentioned above, to embed a micro-strategy within MCTS in the proposed way, the micro-strategy should be computationally efficient. However, in general, many popular graph properties are expensive to compute and instead, in some cases, they may be efficiently approximated or partially computed. For example, the number of winning sets may be exponential with respect to the number of vertices in the graph, however, a suitable micro-strategy may be to efficiently analyse a small randomly selected subset of winning sets. 

\section{Types of Game Simplification}
\label{sec-typesofsimplifications}
Even in seemingly medium-sized graph games, it can be difficult to use simulation to compare the performance of MCTS and other similar strategies; however, the approach we propose is able to produce meaningful comparisons in some of these difficult cases. Using the framework of Section \ref{sec-intro}, we describe the approach which is essentially based upon analysing simplified versions of the games. Recall from Section \ref{sec:mbgames} that we are interested in sets of MB games which represent simplifications that change the characteristics of the game in a minor way. For some games and micro-strategies, relatively good performance in the simplified versions may imply good performance in the complex game. If this is the case, then the simplified versions can be used to indirectly search for good strategies to use in the complex game, because comparative simulations are more efficient by design. To investigate this approach, we consider three types of simplifications which are detailed below along with the instances of MB games used in the upcoming experiments. The viewpoint used for the experiments is that Maker is attempting to use simulation in order to improve upon their traditional MCTS strategy. The instances have been chosen so that simulation and comparisons are computationally tractable but become difficult and time-consuming for the more complex instances.

The first type of simplification comes from varying the game board. In this, the winning sets are implicitly given by subraphs which are isomorphic to a given fixed graph, and therefore changing the board (potentially) also changes the winning sets. Different winning sets are essential for this framework, otherwise, the game change becomes uninteresting. The simplifications arise from the small graphs which should contain fewer winning sets and facilitate more efficient simulation. We will examine the following instances, where the simulations involve Maker using a modified MCTS and versing breaker using traditional MCTS
\begin{itemize}
\item Board is a square $k\times k$ grid graph and the winning sets are the vertices of any path of length 7 (7 vertices). The parameter $k$ is varied from 3 to 9.
\item Board is a rectangular $5\times k$ grid graph and the winning sets are all dominating sets of the graph. The parameter $k$ is varied from 3 to 11.
\item Board is an Erd\"{o}s-Renyi random graph\footnote[2]{An Erd\"{o}s-Renyi random graph $\mathcal{G}(n,p)$ has $n$ vertices and each pair of vertices are connected by an edge independently with probability $p$.}  $\mathcal{G}(12,k)$ and the winning sets are all dominating sets of the graph. The parameter $k$ is varied in small increments from 0.2 to 0.5.

\end{itemize}

The second type of simplification comes from varying the type of winning sets in an MB game, which has an overall effect that is quite similar to the first type of simplification. Again, the winning sets are implicitly given by subraphs that are isomorphic to a given fixed graph, for example, all paths of length $k$. The winning sets are then varied by altering the parameter $k$ and simplifications arise from the ``small'' values of $k$ in which it should be easy for a player to win and it is also computationally easy to identify the win. We will examine the following instances, where the simulations involve Maker using a modified MCTS and versing breaker using traditional MCTS
\begin{itemize}
\item Board is a square $6\times 6$ grid graph and the winning sets are the vertices of any path of length $k$ ($k$ vertices). The parameter $k$ is varied from 3 to 8.
\item Board is an Erd\"{o}s-Renyi random graph $\mathcal{G}(15,0.2)$ and the winning sets are the vertices of any path of length $k$. The parameter $k$ is varied from 3 to 6.
\end{itemize}

The third type of simplification comes from varying the complexity of the strategies being used. Because strategies can vary greatly in characteristics, it is more difficult to describe how to appropriately change them in a ``minor'' way. However, with our framework, there are some obvious approaches. We will simulate the following instances
\begin{itemize}
\item Maker uses modified MCTS with $k$ rollouts versus Breaker using traditional MCTS, also with $k$ rollouts. The parameter $k$ is varied from 3 to 8.
\item The simplified strategies will be modelled as Maker solely using a micro-strategy, without any MCTS, versus Breaker using a uniform random strategy. The complex strategies will be modelled as Maker using modified MCTS versus Breaker using traditional MCTS.
\end{itemize}
For this simplifiction type, the board and winning sets are considered fixed. For the first instance above, the board is a Flower Snark graph on 28 vertices \cite{flowersnarks} and the winning sets are any path of length 7 (7 vertices). The Flower Snarks were chosen in order to examine a relatively sparse graph which also contains a large amount of symmetry in its structure. In the second instance above, the board is a rectangular $5\times 7$ grid graph and the winning sets are the vertices of any dominating set.

\section{Experiments and Discussion}\label{sec-exp}
The experiments are designed to provide initial results for the broad approach outlined in Sections \ref{sec-intro} and \ref{sec-typesofsimplifications} of using game simplifications to aid in strategy development. For the instances described in Section \ref{sec-typesofsimplifications}, different strategies will be simulated and their performance compared between the different values of the game parameters. We make 29 different strategies by modifying the traditional MCTS using 29 different micro-strategies of types similar to those listed in Section \ref{sec-ms}. In the upcoming results, there are many algorithmic choices and parameter settings involved and we omit discussing these for this initial conceptual exploration. Where possible, each different game was simulated 10,000 times and the results were recorded. For more difficult instances, a timeout was needed on both the individual games and the overall runtime, reducing the total number of simulations for that game. The upper bounds of the parameters that define the game instances were chosen by identifying when the number of timeouts became significant. Although some of these games have known (sometimes obvious) winning strategies, we are interested in their characteristics as the games are varied, and so they are still interesting for an algorithm that does not inherently know a winning strategy.

% From this, it will be seen that by simulating modified MCTS in simplified games, improvements can be found to a standard MCTS for an original larger/complex game. The second investigation will be demonstrated that can be useful when MCTS is not able to be effectively evaluated in neither the original game nor the simplifications. 

\begin{figure}[H]
 \begin{center}
  \begin{tikzpicture}[baseline]
     \begin{axis}[xlabel={$k$},
   ytick={15,30,45,60},ymin=15,
   ymax=60,
 ylabel={Maker win \%},
 mark size=0pt,
 width=\textwidth,
 height=6cm,
 minor y tick num=5,
xtick={3,5,7,9,11},
 label style={font=\scriptsize}, 
 tick label style={font=\scriptsize},
 title={$P_{5} \times P_k$, $W=\{\text{dominating sets}\}$},
xtick pos=left,
 ytick pos=left]
 ]

 \foreach \i in {2,...,25}
 {
 \addplot table [y index=\i,x index=0,col sep = comma,header=false]{c_mctsgd.csv};
 }
 \addplot[line width=2pt] table [y index=1,x index=0, col sep = comma,header=false]{c_mctsgd.csv};
     \end{axis}
 \end{tikzpicture}
 \begin{tikzpicture}[baseline]
     \begin{axis}[xlabel={$k$},
   xtick={4,5,6,7,8,9},
 ylabel={Maker win \%},
 mark size=0pt,
 width=\textwidth,
 height=6cm,
 %minor y tick num=4,
 %minor x tick num=1,
  label style={font=\scriptsize}, 
 tick label style={font=\scriptsize},
 title={$P_{k} \times P_k$, $W=\{7\text{-paths}\}$},
xtick pos=left,
 ytick pos=left]
 ]
 \addplot[line width=2pt] table [y index=1,x index=0, col sep = comma,header=false]{growing_grid_path.csv};
 \foreach \i in {2,...,25}
 {
 \addplot table [y index=\i,x index=0,col sep = comma,header=false]{growing_grid_path.csv};
 }
     \end{axis}
 \end{tikzpicture}
  \begin{tikzpicture}[baseline]
     \begin{axis}[xlabel={$k$},
 ylabel={Maker win \%},
 mark size=0pt,
 width=0.5\textwidth,
 height=6cm,
  label style={font=\scriptsize}, 
 tick label style={font=\scriptsize},
 title={\small $\mathcal{G}(12,k)$, $W=\{\text{dominating sets}\}$},
xtick pos=left,
 ytick pos=left]
 ]
 %\addplot[line width=2pt] table [y index=1,x index=0, col sep = comma,header=false]{data/newmicro2_rd.csv};

  \addplot[line width=2pt] table [y index=1,x index=0, col sep = comma,header=false]{growing_rand_12_dom.csv};
 \foreach \i in {2,...,26}
 {
   %\addplot table [y index=\i,x index=0,col sep = comma,header=false]{data/newmicro2_rd.csv};
 \addplot table [y index=\i,x index=0,col sep = comma,header=false]{growing_rand_12_dom.csv};
  }
     \end{axis}
 \end{tikzpicture}
 \begin{tikzpicture}[baseline]
     \begin{axis}[xlabel={$k$},
 ylabel={Relative to max of parameter},
 mark size=0pt,
 width=0.5\textwidth,
 height=6cm,
  label style={font=\scriptsize}, 
 tick label style={font=\scriptsize},
 title={\small $\mathcal{G}(12,k)$, $W=\{\text{dominating sets}\}$},
xtick pos=left,
 ytick pos=left]
 ]
 %\addplot[line width=2pt] table [y index=1,x index=0, col sep = comma,header=false]{data/newmicro2_rd.csv};

  \addplot[line width=2pt] table [y index=1,x index=0, col sep = comma,header=false]{growing_rand_12_dom_rel.csv};
 \foreach \i in {2,...,26}
 {
   %\addplot table [y index=\i,x index=0,col sep = comma,header=false]{data/newmicro2_rd.csv};
 \addplot table [y index=\i,x index=0,col sep = comma,header=false]{growing_rand_12_dom_rel.csv};
  }
     \end{axis}
 \end{tikzpicture}
 \end{center}
 \caption{Simulations for three sets of MB games, with Maker win percentage across different parameter values. Different lines correspond to different modified MCTS strategies for Maker and Breaker always uses traditional MCTS. Thick lines are traditional MCTS for Maker. Because the bottom left figure is hard to visualise, the bottom right is an alternative way to make the desired observations.\label{fig-simp1}}
 \end{figure}
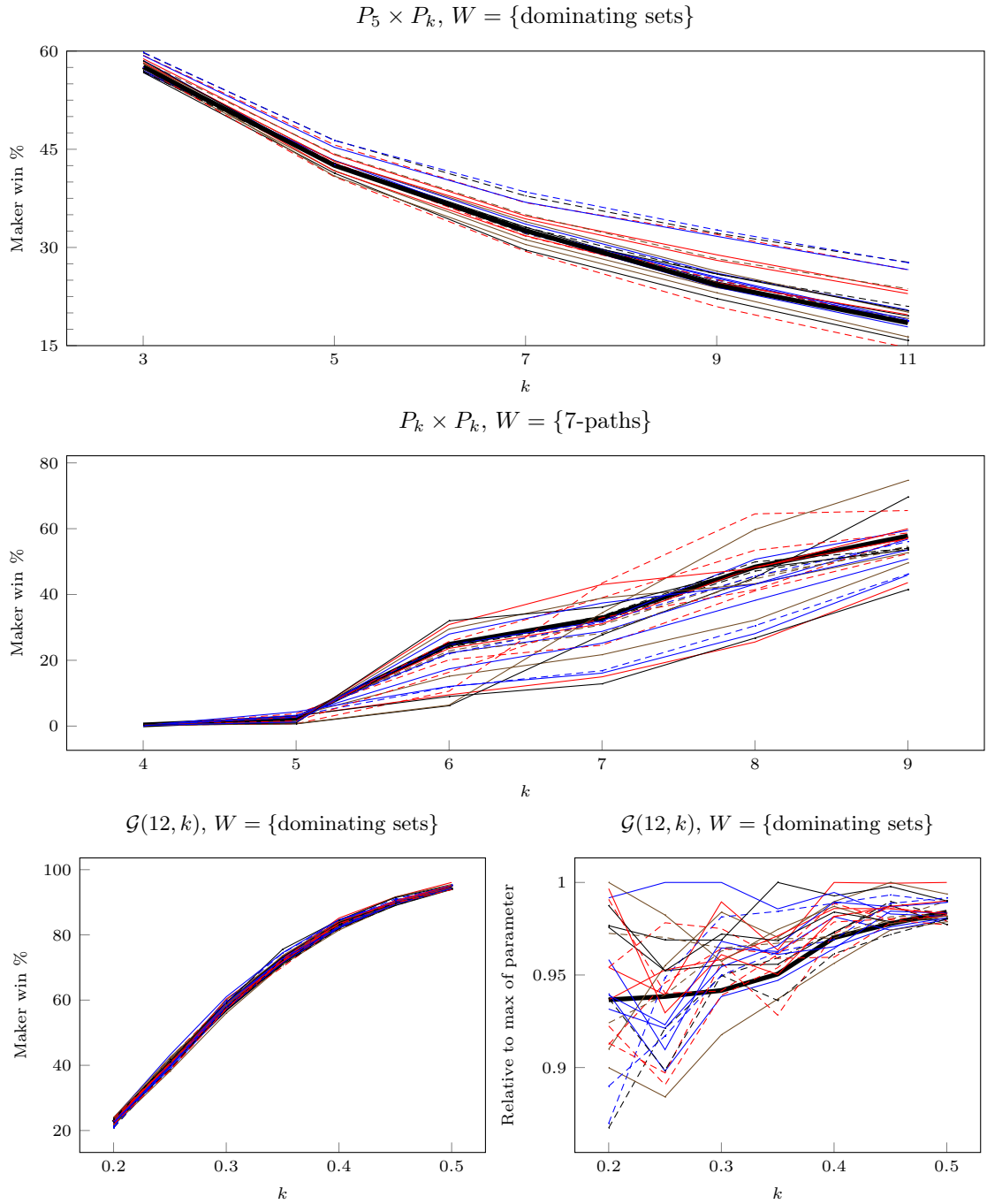

For the first type of simplification described in Section \ref{sec-typesofsimplifications}, the game board is changed, but the type of winning sets is fixed. Maker simulates themselves using modified MCTS and versus Breaker, who uses traditional MCTS. The resulting win percentages for Maker for the different modified MCTS strategies are displayed in Figure \ref{fig-simp1} for the three different sets of MB games. For games on grid graphs, simplified versions retain most of their relative strategy performance as the grid size changes. In these cases, finding a micro-strategy that performs well in the smaller instances has a good chance of providing the desired improvement to Makers traditional MCTS for {\em any} larger sized instance of these games. These instances are examples of games and simplifications in which our approach is successful and an improved strategy can be identified. In the games with 7-paths as the winning sets, there is less stability in the relative performance of the strategies. Indeed, some strategies are significantly worse than traditional MCTS for small instances and then significantly better for larger instances. In this case, it is difficult to concretely identify micro-strategies that provide improvements for the larger instances. However, it is still the case that once a range of smaller instances have been compared, some micro-strategies can be identified as potentials for improvement, which can then be individually analysed. In the random graphs, the win percentages vary greatly over the range of the parameter $k$, making a similar visualisation difficult. Instead, in the bottom right figure, for each parameter value, the win percentage relative to the best strategy is visualised. This means that for each parameter value, the ranking of each strategy is retained between the bottom left and right figures. Again, a few strategies can be identified which consistently perform well and can form the basis of further investigation. However, in general, this is an example where the approach would not be effective without further analysis. The large amount of noise is reasonable in a random graph setting, where there is an enormous number of different games that can possibly be generated. 

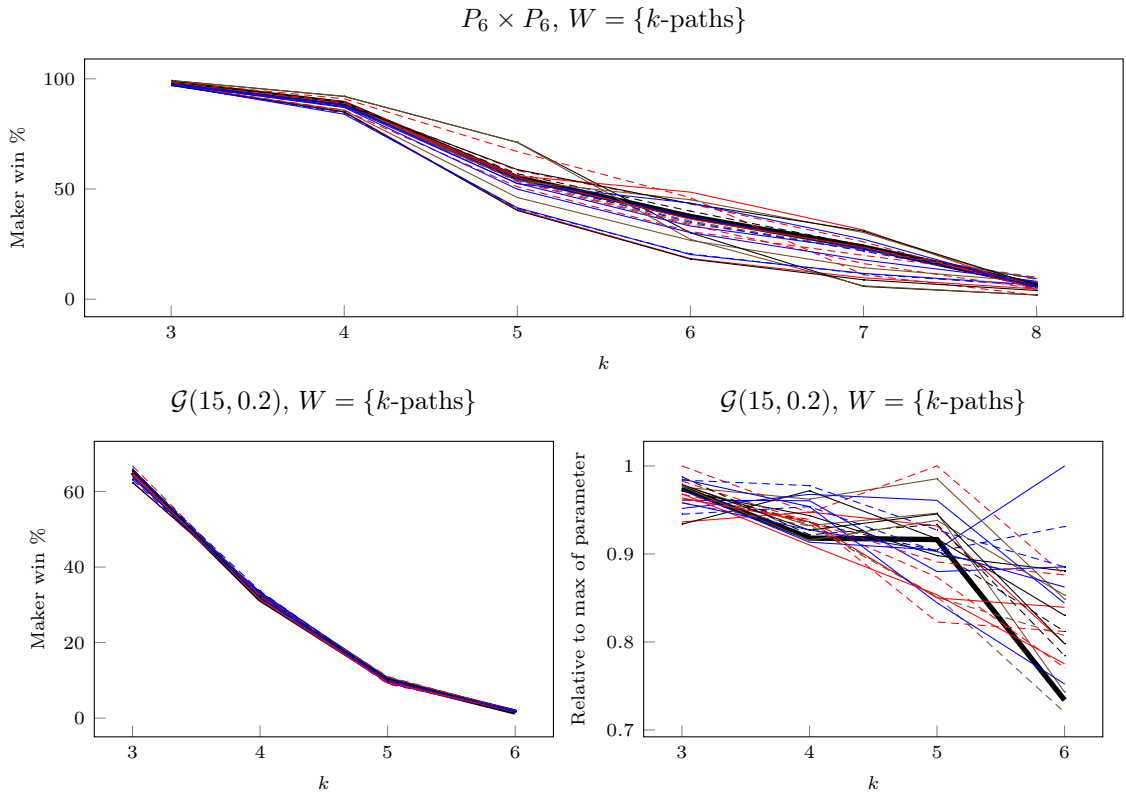
\begin{figure}[H]
 \begin{center}
 \begin{tikzpicture}[baseline]
     \begin{axis}[xlabel={$k$},
   xtick={3,4,5,6,7,8},
 ylabel={Maker win \%},
 mark size=0pt,
 width=\textwidth,
 height=5cm,
 %minor y tick num=4,
 %minor x tick num=1,
  label style={font=\scriptsize}, 
 tick label style={font=\scriptsize},
 title={$P_{6} \times P_6$, $W=\{k\text{-paths}\}$},
xtick pos=left,
 ytick pos=left]
 ]
 \addplot[line width=2pt] table [y index=1,x index=0, col sep = comma,header=false]{growing_path_grid.csv};
 \foreach \i in {2,...,25}
 {
 \addplot table [y index=\i,x index=0,col sep = comma,header=false]{growing_path_grid.csv};
 }
     \end{axis}
 \end{tikzpicture}
 \begin{tikzpicture}[baseline]
     \begin{axis}[xlabel={$k$},
   xtick={3,4,5,6},
 ylabel={Maker win \%},
 mark size=0pt,
 width=0.5\textwidth,
 height=5.5cm,
 %minor y tick num=4,
 %minor x tick num=1,
  label style={font=\scriptsize}, 
 tick label style={font=\scriptsize},
 title={$\mathcal{G}(15,0.2)$, $W=\{k\text{-paths}\}$},
xtick pos=left,
 ytick pos=left]
 ]
 \addplot[line width=2pt] table [y index=1,x index=0, col sep = comma,header=false]{growing_path_rand_15_point_2.csv};
 \foreach \i in {2,...,25}
 {
 \addplot table [y index=\i,x index=0,col sep = comma,header=false]{growing_path_rand_15_point_2.csv};
 }
     \end{axis}
 \end{tikzpicture}
 \begin{tikzpicture}[baseline]
     \begin{axis}[xlabel={$k$},
   xtick={3,4,5,6},
 ylabel={Relative to max of parameter},
 mark size=0pt,
 width=0.5\textwidth,
 height=5.5cm,
 %minor y tick num=4,
 %minor x tick num=1,
 label style={font=\scriptsize}, 
 tick label style={font=\scriptsize},
 title={$\mathcal{G}(15,0.2)$, $W=\{k\text{-paths}\}$},
xtick pos=left,
 ytick pos=left]
 ]
 \addplot[line width=2pt] table [y index=1,x index=0, col sep = comma,header=false]{growing_path_rand_15_point_2_rel.csv};
 \foreach \i in {2,...,25}
 {
 \addplot table [y index=\i,x index=0,col sep = comma,header=false]{growing_path_rand_15_point_2_rel.csv};
 }
     \end{axis}
 \end{tikzpicture}
 \end{center}
 \caption{Simulations for two sets of MB games, with Maker win percentage across different parameter values. Different lines correspond to different modified MCTS strategies for Maker and Breaker always uses traditional MCTS. Thick lines are traditional MCTS for Maker. Because the bottom left figure is hard to visualise, the bottom right is an alternative way to make the desired observations.\label{fig-simp2}}
 \end{figure}

For the second type of simplification described in Section \ref{sec-typesofsimplifications}, the type of winning sets is changed, but the game board is fixed. Again, Maker uses modified MCTS and versus Breaker, who uses traditional MCTS. The resulting win percentages for Maker are shown in Figure \ref{fig-simp2} for the two different sets of MB games. In these simplifications, there is again less stability in the relative performance of the strategies. However, there is still an evident general trend that is visible after simulating across a range of simplifications. In the grid graphs, unfortunately this general trend mostly indicates that a lot of the tested strategies are worse than traditional MCTS for the larger instances. However, there are a few strategies that are identified to be consistently better in the moderate-sized instances. For the largest tested instances, the win rates are small and the performance differences are due to simulation noise. One conclusion is that these good performing strategies may be improvements, not only on these games, but also for games on much larger square-grids with correspondingly sized (i.e. not too large) paths as the winning sets. In the random graphs, similarly to the first simplification type, the win percentages vary greatly making visualisation difficult. Instead, the bottom right figure visualises the win percentage relative to the best strategy for each parameter value. Again, the random graphs provide a significant amount of noise, demonstrating an example where the approach has failed. However, consistently good strategies can still be identified for further analysis.

For the third type of simplification described in Section \ref{sec-typesofsimplifications}, the game board and type of winning sets are fixed, and the two different strategy simplifications are considered. The first strategy simplification is for Maker to use modified MCTS and breaker to use traditional MCTS, but the number of rollouts within the MCTS process is varied, providing a simplification when there are a small amount of rollouts and complex versions from many rollouts. The resulting win percentages for Maker are shown in the top of Figure \ref{fig-simp3}. In these games, the relative performance of the different strategies is mostly retained as the number of rollouts is changed. The best performing strategies can be easily identified after simulating only a few of the smaller (efficient) parameter settings. These instances are another example of games and simplifications in which our approach is successful. The second strategy simplification comes from both players using only a micro-strategy, with no MCTS. Maker considers all of the 29 micro-strategies and versus Breaker who uses only a uniform random micro-strategy. The corresponding complex version of the game is for Maker to use a modified MCTS and versus Breaker who uses traditional MCTS. The resulting win percentages for Maker are shown in the bottom of Figure \ref{fig-simp3}. Although there are only two ``parameters'' in this example, for simplicity, we keep the display style consistent with the others. This instance displays very little correlation between relative strategy performance of the micro-strategies and the corresponding modified MCTS. Unfortunately, this also seemed to be the case for several other tested game boards and winning sets (and various other parameter settings). We predict that modelling Breaker using a uniform random strategy is too large of a simplification and obscures the desired information. Instead, an intermediate complexity strategic simplification should be more effective, such as varying the rollouts in the previous example. 

\begin{figure}[H]
 \begin{center}
 \begin{tikzpicture}[baseline]
     \begin{axis}[xlabel={Number of MCTS rollouts},
 ylabel={Maker win \%},
 mark size=0pt,
 width=\textwidth,
 height=5cm,
 %minor y tick num=4,
 %minor x tick num=1,
   label style={font=\scriptsize}, 
 tick label style={font=\scriptsize},
 title={Flower Snark 7 (28 vertices), $W=\{7\text{-paths}\}$},
xtick pos=left,
 ytick pos=left]
 ]
 \addplot[line width=2pt] table [y index=1,x index=0, col sep = comma,header=false]{growing_rollouts_flower_snark.csv};
 \foreach \i in {2,...,25}
 {
 \addplot table [y index=\i,x index=0,col sep = comma,header=false]{growing_rollouts_flower_snark.csv};
 }
     \end{axis}
 \end{tikzpicture}\\
 \begin{tikzpicture}[baseline]
     \begin{axis}[
   xtick = {0,1},
   xticklabels={micro-strategy,modifed MCTS},
 ylabel={Maker win \%},
 mark size=0pt,
 width=\textwidth,
 height=4.5cm,
 %minor y tick num=4,
 %minor x tick num=1,
   label style={font=\scriptsize}, 
 tick label style={font=\scriptsize},
 title={$P_{5} \times P_7$, $W=\{7\text{-paths}\}$},
xtick pos=left,
 ytick pos=left]
 ]
 \addplot[line width=2pt] table [y index=1,x index=0, col sep = comma,header=false]{growing_strat.csv};
 \foreach \i in {2,...,25}
 {
 \addplot table [y index=\i,x index=0,col sep = comma,header=false]{growing_strat.csv};
 }
     \end{axis}
 \end{tikzpicture}
 \end{center}
 \caption{Two sets of MB games, with Maker win percentage across different parameter values. Different lines correspond to different modified MCTS strategies for Maker and Breaker always uses traditional MCTS. Thick lines are traditional MCTS for Maker.\label{fig-simp3}}
 \end{figure}

\section{Conclusion}
We have experimentally examined some methods for simplifying games in which relative strategy performance can be retained between simplified and complex versions of the games. Some instances clearly displayed this property, such as the dominating sets in the rectangular grid graphs and also the varying number of rollouts in MCTS. Other instances exhibited much more noise in strategy performance across different parameter values, such as in the random graphs. However, even in some of the noisy cases, consistently good strategies could still be identified, which at least facilitates further examination. When good strategies can be identified from the simplifications, it immediately leads to our method for finding improvements to a strategy in the original, complex version of the game. For general games, the approach can be summarised as follows 

\begin{itemize}
\item Identify ways to simplify the game in which some characteristics of the original game are retained. 
\item Analyse micro-strategies and modified MCTS in the simplified versions.
\item Develop an ensemble strategy for the original game, from the analysis in the simplified games.
\end{itemize}

For this approach, it makes sense to have a large set of micro-strategies on-hand that can be applied to a wide variety of games in order to do a sweeping exploratory analysis. Then, high-performing micro-strategies can be identified and analysed in the current game setting, possibly leading to further improvements. The observed performance relationships are dependent on both the strategies and the games involved, and in other experiments we have observed games where there is essentially no such relationship \cite{haythorpe}. Simply speaking, this occurred when incrementing a game parameter tended to drastically alter the game. 

% One deficiency of MCTS arises when the size of the search tree far outweighs the practical number of iterations of MCTS. In these cases, MCTS rarely explores similar game states and essentially degenerates into random play. Our approach can also be used as a preliminary investigation into strategising in games such as these where the simplified strategies and games are used to provide hueristic evidence of strategy quality.

The modified MCTS could be replaced by other advanced strategies, specifically, any which can integrate with a micro-strategy. It is of interest to better understand the instances in which such an approach is successful. In the simplest case, there are examples of games where copies of small games can be pasted together, and winning strategies can be naturally extended to the combined game. Our results show that other ways of combining games can also yield strong strategies, at least heuristically.

\end{document}